\documentclass[aps,pra,twocolumn,showpacs,10pt]{revtex4-2}
\usepackage{graphicx}
\usepackage{amsmath}
\usepackage{amsfonts}
\usepackage{verbatim}
\usepackage{bm}
\usepackage{graphicx}
\usepackage{subfigure}
\usepackage{epstopdf}
\usepackage{siunitx}
\usepackage{braket}
\usepackage{tabularx}
\usepackage{blindtext}
\usepackage{amsmath}
\usepackage{verbatim} 
\usepackage[version=3]{mhchem} 
\usepackage[symbol]{footmisc}
\usepackage{gensymb}
\usepackage{xfrac}
\usepackage[]{algorithm2e}
\usepackage[left=2.5cm,right=3cm,top=2cm,bottom=2cm]{geometry}
\begin{document}

\title{Compressed sensing enabled high-bandwidth and large dynamic range magnetic sensing}

\author{Galya Haim} 
\affiliation{Institute of Applied Physics, Hebrew University, Jerusalem 91904, Israel}
\affiliation{School of Physics, The University of Melbourne, Parkville, Victoria 3010, Australia }

\author{Chris Mullarkey}
\affiliation{University of Rochester Department of Physics and Astronomy}
\affiliation{Racah Institute of Physics, Hebrew University, Jerusalem 91904, Israel}

\author{John Howell} 
\affiliation{Institute for Quantum Studies, Chapman University, Orange, CA 92866, USA}
\affiliation{Racah Institute of Physics, Hebrew University, Jerusalem 91904, Israel}

\author{Nir Bar-Gill} 
\affiliation{Institute of Applied Physics, Hebrew University, Jerusalem 91904, Israel}
\affiliation{Racah Institute of Physics, Hebrew University, Jerusalem 91904, Israel}

\date{\today}

\begin{abstract}

Electron Spin Resonance (ESR) is a widely common method in the field of quantum sensing. Specifically with the Nitrogen-Vacancy (NV) center in diamond, used for sensing magnetic and electric fields, strain and temperature.
However, ESR measurements are limited in temporal resolution, primarily due to the large number of data points required—especially in high dynamic range regimes—and the need for extensive averaging caused by low signal-to-noise ratio (SNR).
This study introduces a novel application of compressed sensing (CS) for magnetic sensing using NV centers. By comparing CS with conventional raster scanning, we demonstrate the potential of CS to enhance sensing applications. Experimental results, supported by simulations, show an improvement of factor 3 in measurement accuracy in low SNR data, which also translates to achieving the same accuracy with only $15\%$ of the data points.
Moreover, the proposed approach is not confined to NV centers but can be extended to ESR measurements in other systems, broadening its applicability in quantum sensing.
\end{abstract}

\maketitle

\section{Introduction}
Compressed Sensing (CS) has been established as a powerful technique, allowing for the accurate reconstruction of signals with significantly fewer measurements than traditional methods in a broad range of applications \cite{5466604, samplingTheory, csNappli}. 
This method leverages the principle that a signal, if sparse in a particular basis, can be effectively reconstructed from a limited number of measurements, provided these measurements are orthogonal or at least incoherent with respect to the sparsifying basis. By focusing on solutions that maximize sparsity, CS ensures high fidelity reconstruction even in noisy environments \cite{Candes2008-sz,Duarte2011-fc}. The technique's ability to reduce measurement time, cost, and complexity has led to its widespread application across diverse fields including bio-sensing \cite{Craven2015-lw}, Magnetic Resonance Imaging \cite{Jaspan2015-ao}, and LIDAR \cite{Howland2013-kz, Sher2019-tv}Additionally, CS is making
significant strides in quantum sensing and state tomography \cite{PhysRevLett.105.150401}.

Standard formulations of CS and their corresponding proofs of accuracy and convergence often assume that the sparse-representation of an underlying signal is in an orthonormal basis, for instance in a Fourier or wavelet basis. Recently increased attention is being paid to CS techniques where the sparse representation is neither normalized or orthogonal, and may even be highly overcomplete \cite{Sadeghi2014-jf, Candes2011-xv}. \par

Here we present CS for quantum sensing of magnetic fields based on nitrogen-vacancy (NV) centers in diamond \cite{Pham2013-gy}, using a highly overcomplete sparse basis in the frequency space. We demonstrate that a CS - based measurement benefits both in terms of measurement time and accuracy in low SNR data, where CS achieved the same error as raster scanning with as few as $15\%$ of the data points, and a lower error overall by a factor 2 and up to 5.

NV magnetic sensing has been established as a useful and diverse technique \cite{RevModPhys.92.015004,Meirzada2021, Ninio2021-yu, Grant2023, McCloskey2022, Huxter2022,Li2023}, leveraging the robustness of the diamond sensor, along with the high-resolution and high sensitivity afforded by the optical readout of the signal and the favorable coherence properties of the NV defects \cite{Pham2013-gy, DOHERTY20131, RevModPhys.92.015004}.

\begin{figure*}[tbh]
\includegraphics[width=1\textwidth] 
{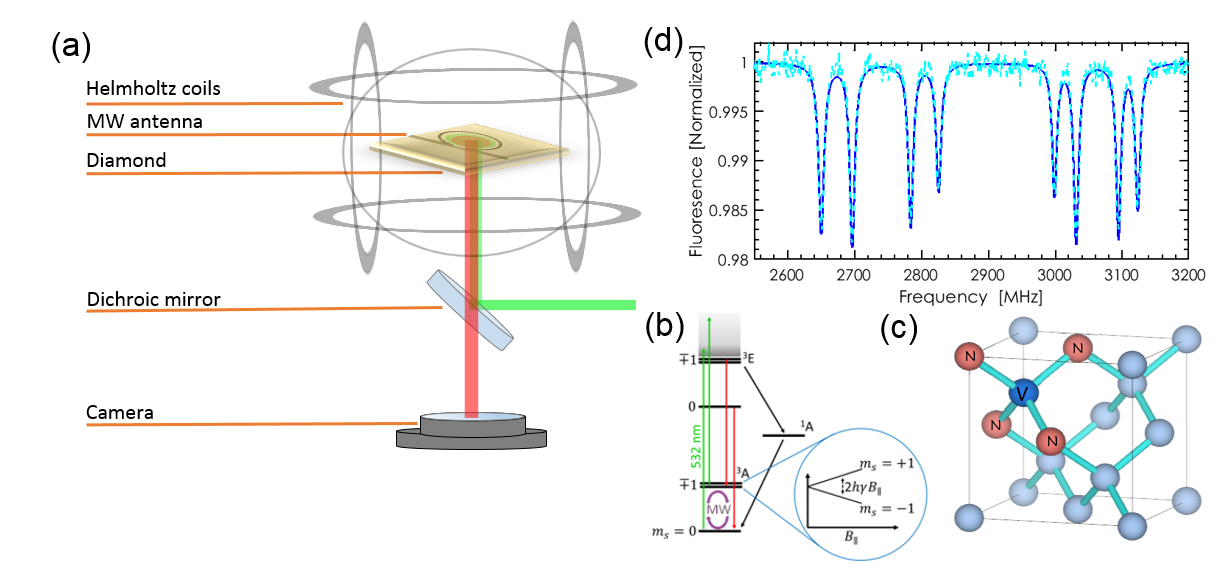}
\caption{\label{fig:NV_setup} (a) Experimental setup schematic, NV's are excited with green laser, manipulated with MW, and red fluorescence is imaged onto a sCMOS camera, in an Epi-illumination scheme. Helmholtz coils generate a magnetic field to distinguish between the resonances of the different orientations.
(b) NV center's electron energy levels, ground state degeneracy is lifted due to Zeeman shift.
(c) Four possible NV orientations in the diamond lattice.
(d) ESR spectrum of NV ensemble, 8 resonances correspond to \(m_s = \pm 1\) of each orientation. }
\end{figure*}

 Standard NV magnetic sensing and imaging techniques are predicated on the accurate determination of ground-state spin energy levels positions (figure \ref{fig:NV_setup} (b)). This is achieved through Electron Spin Resonance (ESR) measurement (figure \ref{fig:NV_setup} (d)) \cite{RevModPhys.92.015004, DOHERTY20131, 10.1063/1.3337096, Clevenson2015}. The spin state of the NV center is optically initialized, followed by the detection of fluorescence while varying microwave frequencies. Resonances are identified by a decrease in recorded fluorescence counts, as the NV center’s fluorescence is dependent on its spin state. For a detailed description of NV ESR and the experimental setup, please refer to the methods section.

Although this method is effective and well-established, it can be time-consuming to sample a large number of microwave frequency points, along with additional averaging, to accurately identify resonance locations with high precision \cite{RevModPhys.92.015004}. This challenge is particularly pronounced in high magnetic field environments that require a broad frequency range, leading to increased measurement time. This issue is well recognized in the field of high dynamic range magnetic sensing \cite{RevModPhys.92.015004, 10.3389/fncom.2013.00137, McCloskey2022}, where extensive sampling can result in diminished sensitivity, and bandwidth. Addressing this challenge is crucial for enhancing the efficiency and accuracy of magnetic sensing.

We note that in most scenarios and in the values quoted for magnetic sensitivity in the literature, it is assumed that a small signal regime is relevant, such that the spectral locations of the resonances are roughly known, and only measuring small deviations from them is required, such that only a couple of measurements on the slope of each resonance are needed. 
Here we focus on the large signal regime, which is relevant for unknown fields with potentially non-negligible gradients, and for which the standard approach inevitably leads to large frequency sweeps that translate into the limitations mentioned above. In such cases, a CS approach can significantly enhance magnetic sensing.

\section{main}
A conventional ESR raster scan involves sweeping across a complete frequency window, determined by a bias magnetic field and the desired dynamic range, along with averaging to achieve the desired SNR, and minimize the measurement's error. After the measurement is complete, the data is analyzed by fitting the results, identifying the resonance locations, and calculating the vectorial magnetic field based on these resonance frequencies \cite{10.1063/1.3337096}.

We introduced the CS algorithm (Alg. \ref{alg:1}) into the NV magnetic sensing system as follows: Initially, a limited number of randomly selected frequency points are measured. This preliminary phase serves both to establish a baseline for fluorescence counts and their deviations, and to expedite the process, as the algorithm requires some initial data to generate meaningful results. Once this stage is concluded, the measurement protocol shifts to operate within a ``while" loop.

\begin{algorithm}
 \KwResult{Estimate of where Lorentzian Peaks Occur}
 Measure user-defined (small) number of initial projections at randomly chosen frequencies\;
 Get initial estimate of mean reference power and peak locations\;
 \While{fit has NOT converged}{
  choose random frequencies within window given frequency window\;
  measure a new projection at those frequencies\;
  get new reconstruction estimate and peak locations\;
  \If{8 peaks have converged}{
            update basis matrix \textbf{L} at the converged locations\;
            get new reconstruction estimate and peak locations\;
            \If{Reconstruction has converged OR max number of measurements reached}{
                return current reconstruction as best estimate\;
            }
   }
 }
 \caption{CS measurement scheme}
\label{alg:1}
\end{algorithm}

The loop terminates based on one of two conditions: either the result has converged, or a maximum number of points have been measured. The procedure within the loop is as follows: a random frequency point is measured, and the CS algorithm reconstructs the data. If eight Lorentzians are identified — indicating a successful measurement — the peak locations and width values are recorded, and the next data point is measured. The loop concludes when four consecutive successful measurements are achieved, with peak values falling within 2 MHz of each other (See Alg. \ref{alg:1}).
The stopping condition can be adjusted by the user to accommodate the system's parameters and desired result. The number of peaks can also be easily changed for cases in which not all eight Lorenzians are required desired.  

In line with common practices in CS, experiments and simulations were conducted with up to 3 simultaneous frequencies. Measuring multiple frequencies simultaneously in CS improves efficiency and data quality, as it allows for concurrent sampling of different frequency points, thereby enhancing the overall sensitivity and reducing measurement time. 
In contrast, for raster scans, simultaneous measurement of multiple frequencies effectively reduces SNR. Given that raster scanning is highly sensitive to SNR variations, multi-frequency measurements were not employed in raster scans. 

\begin{figure}[tbh]
    \centering
    \subfigure[]{\includegraphics[width=1\linewidth]
    {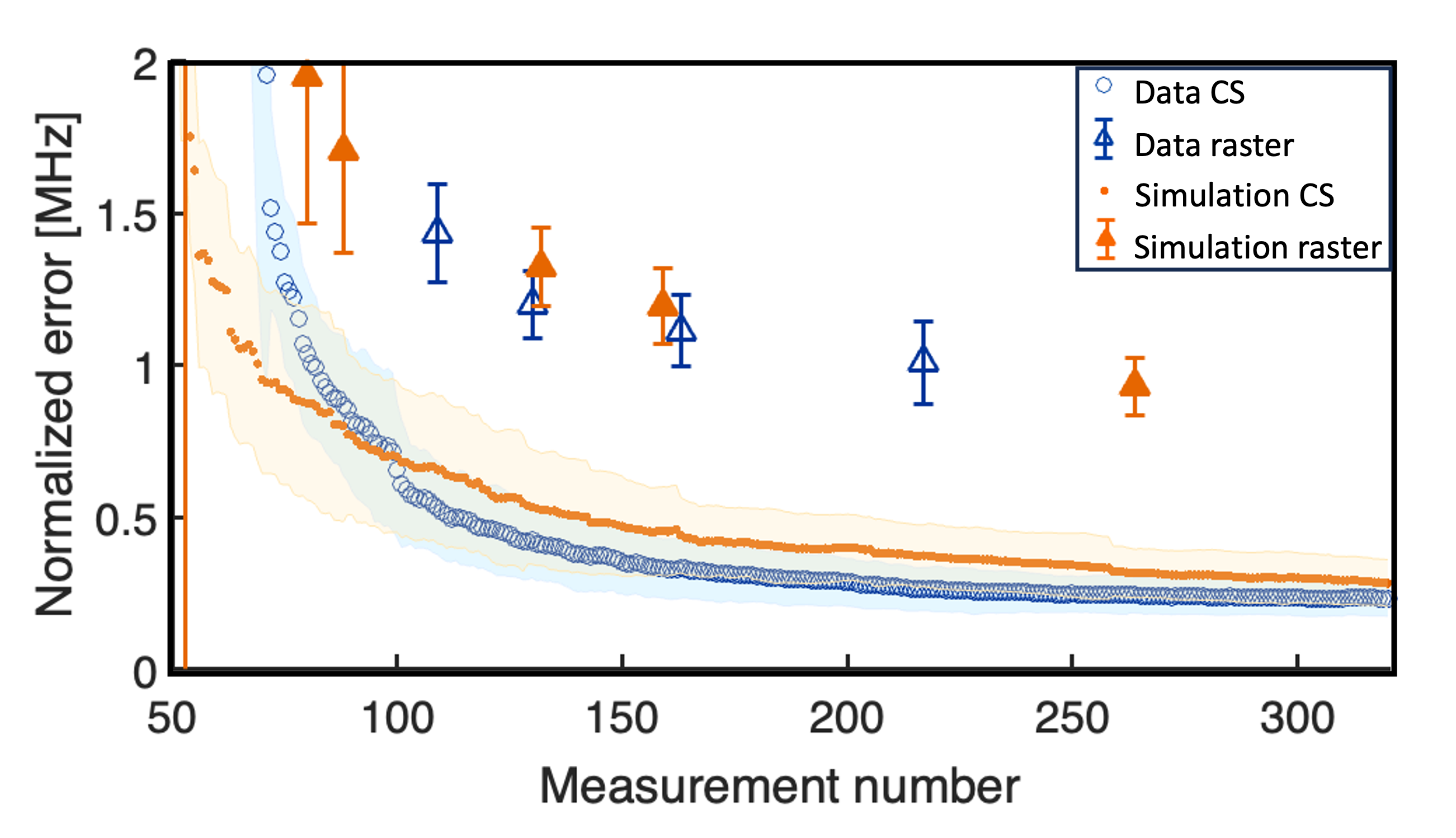}}
    \subfigure[]{\includegraphics[width=1\linewidth]
    {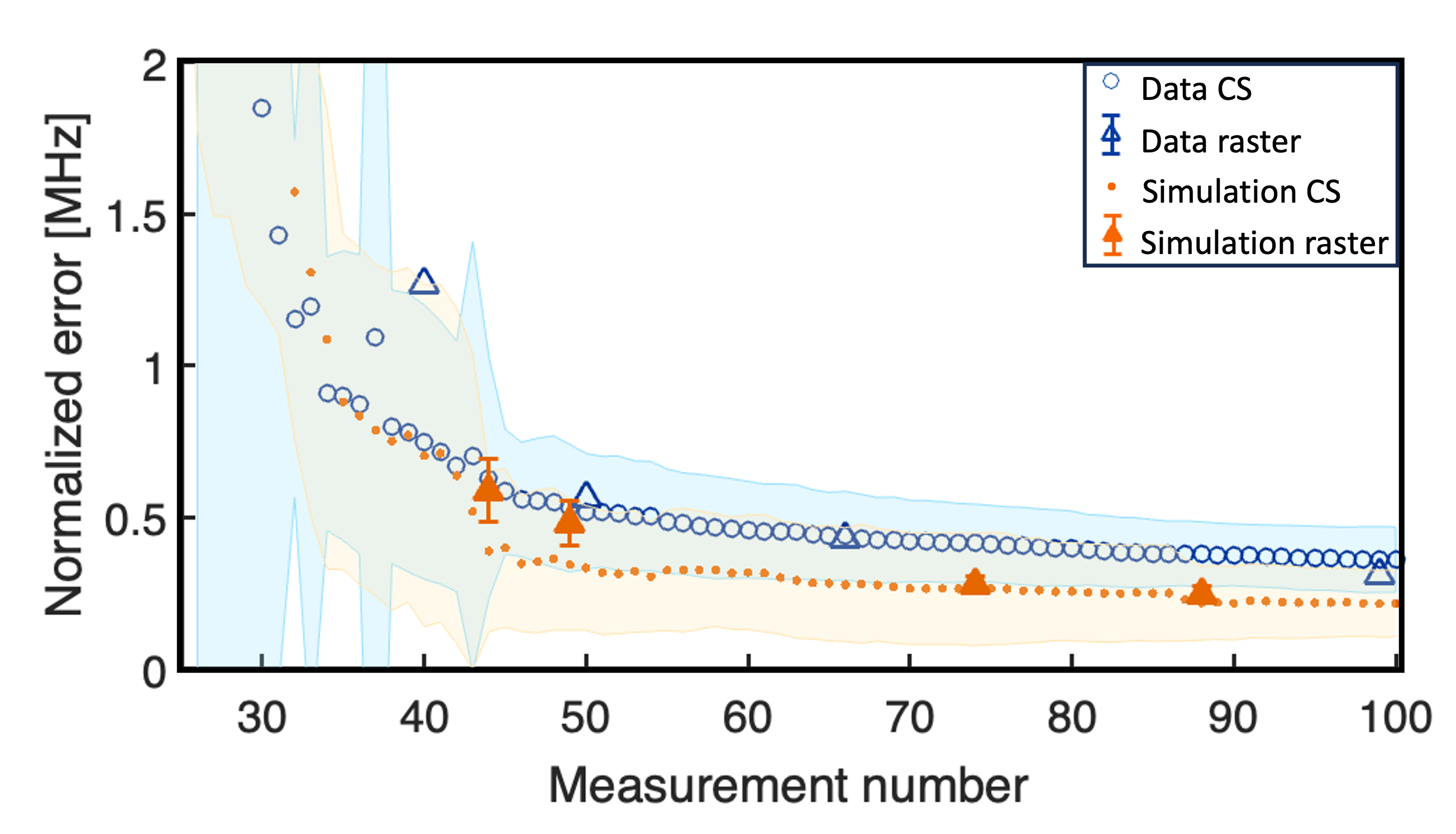}}
    \caption{Experimental data (blue) and simulation results (orange), depicting the averaged normalized error as a function of the number of measurements for a CS protocol (circles), and raster scan (triangles), standard deviation is depicted as shaded regions for CS and error bars for raster.
    We compare measurement results (blue) and simulation results (orange). 
    (a) Simulation and data in high field ($100$ [G]) which corresponds to a frequency window of $650$ MHz. The peak Lorentzian width was $15$ MHz, and the measurement SNR was 3. Simulations generated 57 data samples on which CS and raster scans ran. Real data was acquired on a confocal setup (see methods), where 100 consecutive CS measurements ran, interlaced with a full raster scan every 20 CS measurements (6 in total). The raster scans were sub-sampled in post processing.  
    (b) Simulation and data in low field ($50$ [G]) which corresponds to a frequency window of $400$ MHz. The peak Lorentzian width was $10$ [MHz], and the measurement SNR was 13. Simulations generated 57 data samples on which CS and raster scans ran. Real data was acquired on a wide field setup (see methods), where 100 consecutive CS measurements ran, and one full raster scan that was sub-sampled in post processing.}
    \label{fig:fields}
\end{figure}

We conducted experiments under a relatively high bias magnetic field of approximately 100 Gauss [blue circles in Figure \ref{fig:fields}(a)] and a low bias field of around 50 Gauss [blue circles in Figure \ref{fig:fields} (b)]. These results were compared with simulations (further details provided below) and with raster scans performed before and/or after the CS measurements. The raster scan results were sub-sampled and are represented as triangles in Figures \ref{fig:fields}(a) and \ref{fig:fields}(b).

Although the CS algorithm is intended to terminate measurements once successful results are obtained, data collection was extended until the maximum number of measurements was reached. This was done to collect statistics, characterize and evaluate the method.

The figure of merit used to evaluate the results is the normalized error (for further details, see the methods section), which is defined as $\delta \nu /  \sqrt{P}$. Where P denotes the success probability of peak location identification, and $\delta \nu$ represents the mean absolute error between peak locations. The error is derived from "ground truth" data obtained with a complete raster scan. Compared to those extracted from samples with fewer data points. Both raster scan and CS results were normalized according to the success probability for each number of points to account for the statistical variability inherent in both methods. 
This analysis is useful as it provides a fair comparison between CS and raster scanning, it demonstrates the achievable optimum in terms of the compromise between the number of measurements (and thus measurement time) and the desired sensitivity which is proportional to the normalized error as detailed in the methods section.
 
Simulations were conducted in the same manner as regular measurements, with the sole difference being that the data was derived from simulated data samples. This simulated data was generated using a simplified NV Hamiltonian, which only considered the Zeeman shift. Although the SNR and Lorentzian widths were adjusted to approximate real measured data, there are minor discrepancies, for instance variations in SNR during measurements and differences in Lorentzian widths between peaks and across measurements.
However, as illustrated in both figures \ref{fig:fields}(a) and \ref{fig:fields}(b), the simulations (orange) closely agree with the experimental data (blue), and the trends observed in raster scans and CS are consistent between the two. 

Figures \ref{fig:fields}(a) and \ref{fig:fields}(b) depict experimental and simulation results in high and low field, respectively.
High field experimental data was measured in a confocal microscope (see the methods section for more details on the experimental setup), a total of 100 CS measurements were performed, with a full raster scan conducted after every 20 CS measurements, resulting in 6 raster scans in total. Experimental results were not integrated, each frequency point was measured just once, therefore the SNR was relatively low, approximately 3, and the Lorentzian width was about 15 MHz.
The mean error of CS experiments is depicted in blue circles, and the standard deviation in a blue shaded region. Similarly for simulations, where orange dots plot the mean error and the standard deviation in an orange region. Raster scan's mean error and standard deviation are reported as triangles and error bars, with solid orange triangles for simulations and open blue triangles for experiments. 

Both the data and simulations reveal a substantial advantage of the CS approach over raster sub-sampling, showing a consistent factor of improvement of at least 2. The CS's normalized error is remarkably below 0.5 MHz, even with as few as 100 measurements in a 650 MHz window. 
The normalized error consistently decreases with increasing measurement points for both CS and raster scans, and reaches to a minimal value of 0.5 MHz for raster scans, and 0.1 MHz for CS, at 650 data points. 
This result highlights CS's superiority in low SNR conditions, demonstrating that with approximately 100 data points, CS achieves a normalized error that matches the error obtained by raster scanning with a full frequency sweep. Further details on the impact of SNR on performance are discussed below.

Low-field experimental data [Fig. \ref{fig:fields}(b)] were measured in a wide field microscope (further details are available in the methods section), 100 CS measurements were acquired, along with one full raster scan. The measured SNR was 13 and the peak Lorentzian width was 10 MHz. 
In this case, CS does not offer a significant advantage over raster scanning. 
Raster and CS achieve a comparable normalized error both in real data and in simulations, with CS showing a small improvement under 50 measurements. 
Notably, the CS result with only 40 measurements, approximately 10 percent of the full the raster data set, still achieved an error below 1 MHz.

The difference in CS performance between low and high field arises mostly from the difference in SNR.
The low field data was collected in a wide field setup, and even though each frequency point was measured only once, the data was integrated over multiple pixels. In this case, this integration was achieved without requiring additional measurement time. However, typically, in wide field setups, data is extracted on a per-pixel basis, which means the high SNR observed here is unusual for a single-shot measurement.
 
\begin{figure}[tbh]
\centering
\includegraphics[ width=0.95\linewidth]
{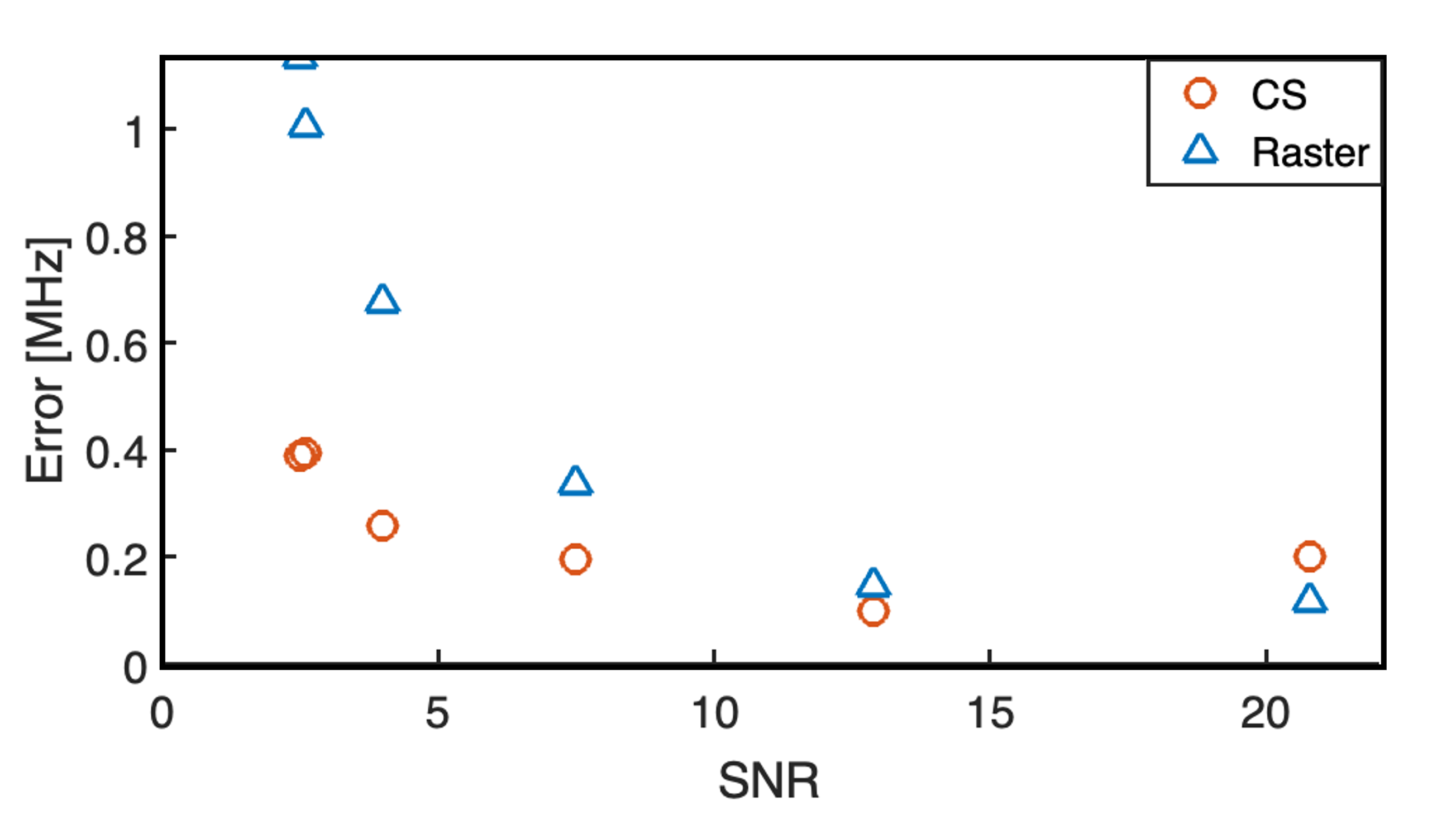}
\caption{(a) Simulated CS (orange circles) and raster's (blue triangle) averaged error for increasing SNR values.
 Each data point is the result of a simulation that ran over 57 data samples in a high field (100 G), and constant Lorentzian width of 10 MHz, from which the averaged error for 217 measurement points is plotted.
}
\label{fig:snr} 
\end{figure}

CS theorem is designed to handle Gaussian noise effectively, which means the algorithm is less sensitive to noise variations compared to raster scanning \cite{Candes2008-sz}.
Figure \ref{fig:snr} depicts simulation results showing error versus SNR across 217 frequency points, averaged over a specific Lorentzian width (10 [MHz]).
At high SNR levels, raster scans exhibit slightly lower error than CS, however, as the SNR decreases, raster scan error increases more sharply, while CS maintains a more consistent performance. With an impressive result for SNR values below 5, CS significantly outperforms raster scanning, achieving an error that is at minimum more than twice as good.

\begin{figure}[tbh]
    \centering
    \includegraphics[ width=0.95\linewidth]
    {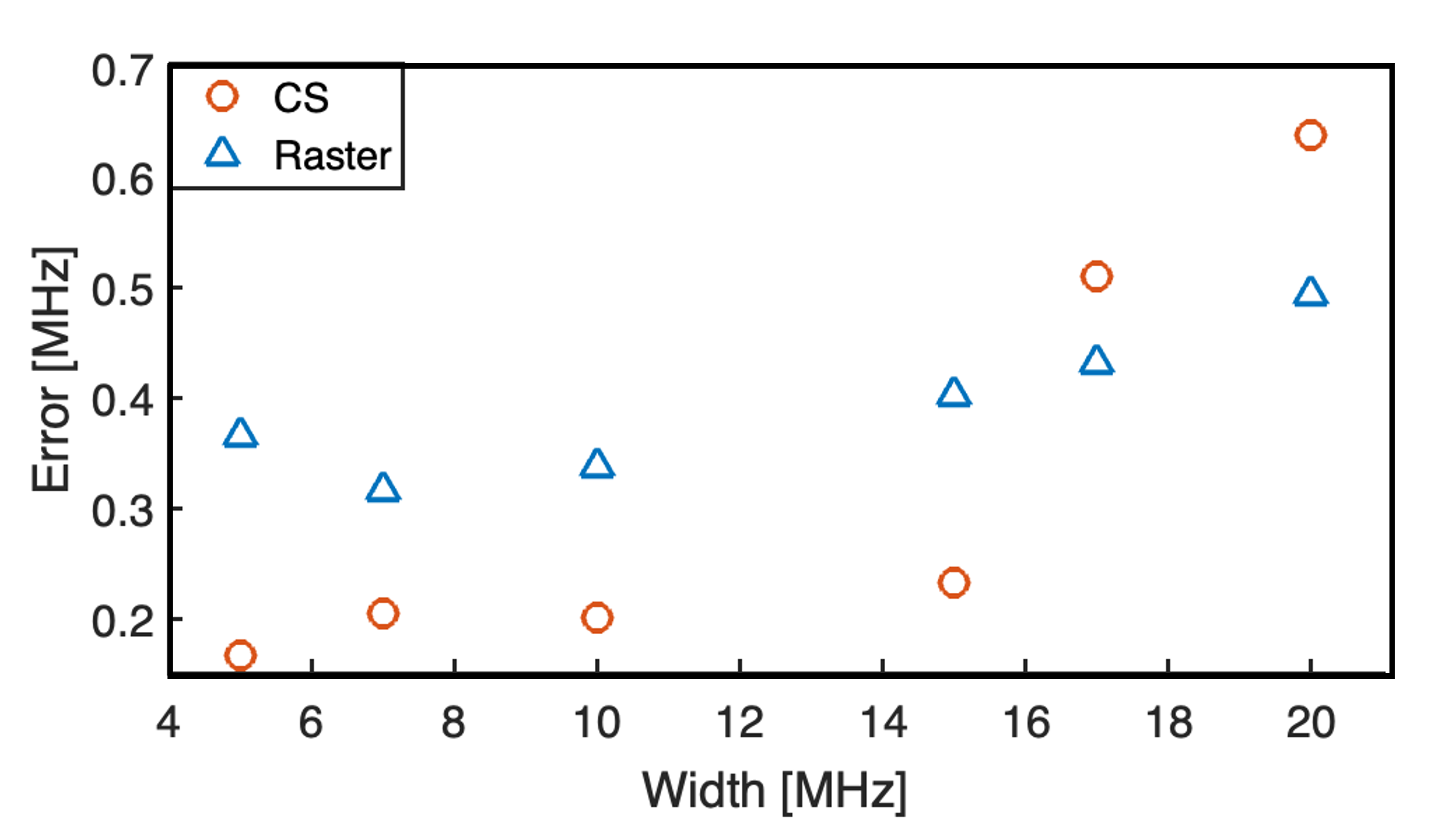}
    \caption{ Simulated averaged error as a function of Lorentzian width. CS (circles) and raster (triangles), both with 217 measurement points, in a bias field of 100 Gauss, Constant SNR of about 7 in all 57 data samples, each one with different field angles, i.e. different resonance frequencies within the frequency window.}
    \label{fig:width}
\end{figure}

 Figure \ref{fig:width} depicts a cross - section comparison of high field simulation results for CS (circles) and raster (triangles) across various Lorentzian widths. The results are averaged over different field angles at a given SNR (9). 
The raster scanning results show a gradual change in error as the Lorentzian width varies. CS results show a relatively consistent error for Lorentzian widths below 15 MHz, with a noticeable increase above this value, which eliminates the advantage of the CS in this regime, for this SNR.  

\begin{figure}[tbh]
\centering
\includegraphics[width=0.95 \linewidth]
{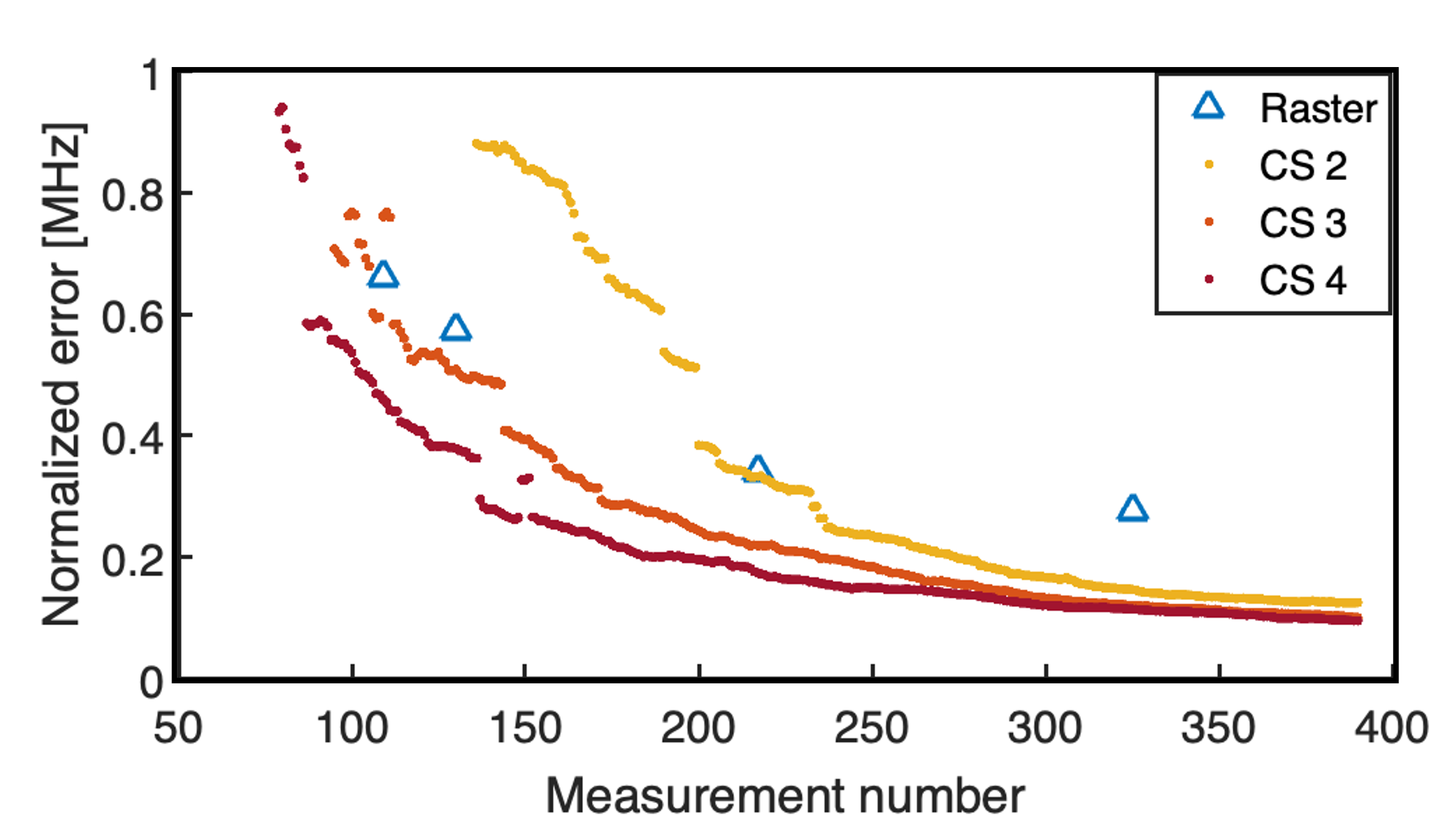}
\caption{Simulated average normalized error in raster scans (blue triangles) and CS with 2,3 and 4 (yellow, orange and red, respectively) simultaneous frequencies. Simulations in high field (100 G), Lorentzian width of 10 MHz, and SNR of about 7.}
\label{fig:freqs} 
\end{figure}

Finally, we examined the impact of measuring simultaneous frequencies, a common practice in CS to improve measurement efficiency.
 Figure \ref{fig:freqs} presents simulation results, comparing raster scan error with CS error across for 2, 3, and 4 simultaneous frequencies.
The data reveals that CS performance significantly improves as the number of simultaneous frequencies increases, particularly for fewer measurements. For example, at 150 data points, using 4 simultaneous frequencies reduces the error by a factor 4 compared to the error observed with just 2 simultaneous frequencies. This enhancement demonstrates the advantage of parallel frequency measurements in CS, allowing for more efficient data acquisition.
All simulation and experimental results presented in this paper utilized 3 simultaneous frequencies. This choice was a balance between performance enhancement and practical implementation in the experimental setup. 
 

\section{Discussion and Conclusion}
In this paper we introduce compressed sensing to enhance NV-based magnetic sensing using the standard ESR technique (for DC fields). The comparison between CS and regular raster scanning of the microwave frequencies reveals CS's clear advantages, particularly in low SNR and large dynamic range environments. Notably, CS achieves a normalized error of approximately 0.5 MHz with around 100 measurements—comparable to the error of raster scanning across a full frequency sweep, depicting an improvement of over a factor of 2. This performance illustrates CS's efficiency in maintaining accuracy with significantly fewer data points, making it particularly valuable in high-bandwidth, high-dynamic range scenarios.

The experimental and simulation results confirm that CS consistently outperforms raster scanning in low SNR conditions, demonstrating a notable factor of improvement in normalized error. CS's effectiveness is further highlighted when measuring multiple simultaneous frequencies, enhancing data acquisition efficiency and accuracy.

In general, CS proves to be a powerful tool for magnetic sensing, offering substantial improvements in both measurement efficiency and accuracy. 
Importantly, we note that the presented approach could significantly impact magnetic sensing applications based on NVs (such as mapping and imaging of field variations in condensed matter samples, biological samples, electrical chip analysis and geophysical surveys). Moreover, it is broadly applicable beyond the NV community, as it can be directly adapted to other systems relying on resonant feature identification, such as various magnetic resonance systems. Finally, we note that future work could address optimized algorithms, e.g., associated with magnetic imaging, whereby each pixel is not analyzed independently, but rather informed by results from previous pixels (similar to techniques used in image-by-image analysis in videos).

\section{Methods}
\subsection{Nitrogen Vacancy center}
The NV center is a point defect in the diamond lattice \cite{DOHERTY20131}, consists of a Nitrogen atom and a vacancy in an adjacent lattice site. In the lattice, the NV can take one of four [1 1 1]  orientations (Fig\ref{fig:NV_setup} c).  

In the presence of an external magnetic field, the the ground state levels \(m_s = \pm 1\) are shifted by the Zeeman effect, proportionally to the field component parallel to the NV axis \ref{fig:NV_setup} (b). 
Finding the new positions of the energy levels is done by optically initializing the spin state, and scanning microwave frequencies (that resonantly excite the electron spin within the ground triplet state), while recording spin dependant fluorescence. This is the well known ESR measurement (Fig\ref{fig:NV_setup} d) \cite{DOHERTY20131, 10.1063/1.3337096, RevModPhys.92.015004}, from which a quantitative vectorial magnetic field can be extracted \cite{10.1063/1.3337096}.

Experimental data was measured on two home built Epi-Illumination microscopes, in which green laser is used to excite the NV's, and manipulation is done with a MW antenna.
Fig \ref{fig:NV_setup} (a) depicts the wide field microscope that was used to acquire small field data. High field data was acquired in a confocal microscope, in which, instead of a camera, fluerescence is collected into a single photon counter. 

The ability to measure magnetic fields is relying on the ability of determining the resonance frequencies accurately. A full ESR measurement would entail measuring linearly spaced MW points, and averaging to improve SNR.
Sensitivity is defined as $\eta = \delta B \sqrt{T}$ where $\delta B$ is the error on magnetic field, and T is the total measurement time. Using the Larmor equation, the sensitivity can be written as $\eta = \sfrac{1}{\gamma}  \delta \nu   \sqrt{T}$, where $\gamma = 2.87 \  \text{MHz}/\text{Gauss}$ is the gyromagnetic ratio of the NV center \cite{Pham2013-gy}. 
To include the probability of success of both CS and raster scanning, we define the normalized error to be: 
$\delta \nu /  \sqrt{P}$ where P is the success probability. A measurement is considered successful when it retrieves 8 Lorentzians. 

\subsection{Compressed Sensing}
In our experiment, a bias magnetic field determines the frequency window over which we expect to find resonances. Define this set of $M$ frequencies $\{\nu_{j}\}_{j=1}^{M}$ at which we may measure and define the set of frequencies $\{\nu_{k}\}_{k=1}^{N}$ as those locations where we allow for the possibility that a resonance is present. Note that it is not necessary, and in practice not the case that any $\nu_{j}=\nu_{k}$, only that they share the window over which resonances actually occur in our system. We can write the total lineshape $A(\nu)$ at frequency  $\nu_{j}$ as 
\begin{align*}
A(\nu_{j}) &= \left(\frac{(\frac{2}{\pi\sigma_{k}})}{1+\frac{4(\nu_{j}-\nu_{k})^{2}}{\sigma^{2}_{k}}}\right)a_{k} \\
&= L_{jk}a_{k}
\end{align*}

Where we have cast this into matrix form and define $\textbf{L}\dot{=}\{L_{jk}\}$ the normalized lorentzian lineshape centered at a location $\nu_{k}$ with width $\sigma_{k}$ measured at frequency $\nu_{j}$ and $\textbf{a}$ as the vector of amplitudes of the lorentzians at each frequency $\nu_{k}$. Note that we expect only a few of the coefficients (two if we had a single NV center with no noise), in $\textbf{a}$ to be non-zero, i.e. \textbf{a} is \emph{sparse}. \par 
Our procedure is as follows: we first sample at a number of frequencies across \emph{all} $\nu_{j}$, and gain an estimate of the average reference power $\bar{\text{P}}_{r}$.
The choice of which frequencies we apply is saved in a $M$-column sampling matrix, each row corresponds to a single projection with entries corresponding to the random frequencies chosen and zeros if that microwave frequency was not chosen at all.\par 
We run a measurement on the chosen frequencies, and a reference measurement (measurement without MW) and attempt to reconstruct the data, after each reconstruction we check whether 8 peaks were found, if the answer is no, the algorithm continues to another measurement with other random frequencies. Once these peak locations have converged, the data is fitted. After four consecutive measurements found identical (within an error of 3 MHz), and the widths of lorentzians are of a value characterizing the diamond, within error, the code breaks the loop. The pseudo-code for this procedure can be found in Alg. 1. \par
We use total variation minimization, specifically the algorithm provided by the L1 \cite{1306.3331} minimization, to reconstruct our 1D signal at each reconstruction step.
\begin{equation*}
\min_{a} \sum_{i} \|D_{i}a\|_{1} \ \text{s.t.} \ \textbf{a}\geq0 \ \text{and} \ \textbf{SLa}=y
\end{equation*}
Where $D_{i}a$ is the discrete gradient of $\textbf{a}$. \par

\section*{Acknowledgments}
This work was financially supported by the European Union’s Horizon 2020 research and innovation programme under FET-OPEN GA No.~828946–PATHOS. G.H. acknowledges support from the Melbourne research scholarship. N.B. acknowledges financial support by the European Commission’s Horizon Europe Framework Programme under the Research and Innovation Action GA No.~101070546–MUQUABIS. N.B.~also acknowledges financial support by the Carl Zeiss Stiftung (HYMMS wildcard), the Ministry of Science and Technology, Israel, the innovation authority (Project No.~70033), and the ISF (Grants No.~1380/21 and No.~3597/21).

\bibliography{main}

\end{document}